\documentclass[aps,reprint]{revtex4-1}

\usepackage{graphicx}% Include figure files
\usepackage{dcolumn}% Align table columns on decimal point
\usepackage{bm}% bold math
\usepackage{textcomp}
\usepackage[usenames,dvipsnames]{xcolor}

\usepackage{amsmath} 
\usepackage{amssymb} 
\usepackage{amsfonts}

\usepackage{array}

\usepackage{multirow}

\begin{document}

\title{Growth modes of partially fluorinated organic molecules on amorphous silicon dioxide}

\author{Mila Miletic}
\affiliation{Research Group Simulations of Energy Materials, Helmholtz-Zentrum Berlin f\"ur Materialien und Energie, Hahn-Meitner-Platz 1, D-14109 Berlin, Germany}
\affiliation{Institut f{\"u}r Physik, Humboldt-Universit{\"at} zu Berlin, Newtonstr. 15, 12489 Berlin, Germany}
 
\author{Karol Palczynski}
\email{karol.palczynski@helmholtz-berlin.de}
\affiliation{Research Group Simulations of Energy Materials, Helmholtz-Zentrum Berlin f\"ur Materialien und Energie, Hahn-Meitner-Platz 1, D-14109 Berlin, Germany}

\author{Joachim Dzubiella}
\email{joachim.dzubiella@physik.uni-freiburg.de}
\affiliation{Applied Theoretical Physics - Computational Physics, Albert-Ludwigs-Universit\"at Freiburg, Hermann-Herder Stra{\ss}e 3, D-79104 Freiburg, Germany}
\affiliation{Research Group Simulations of Energy Materials, Helmholtz-Zentrum Berlin f\"ur Materialien und Energie, Hahn-Meitner-Platz 1, D-14109 Berlin, Germany}

\date{\today}

\begin{abstract}
We study the influence of fluorination on nucleation and growth of the organic para-sexiphenyl molecule (\textit{p}-6P) on amorphous silicon dioxide ({\it a}-SiO$_2$) by means of atomistically resolved classical molecular dynamics computer simulations.
We use a simulation model that mimics the experimental deposition from the vapor and subsequent self-assembly onto the underlying surface.
Our model reproduces the experimentally observed orientational changes from lying to upright standing configurations of the grown layers.
We demonstrate that the increase in the number of fluorinated groups inside the \textit{p}-6P leads to a smoother, layer-by-layer growth on the {\it a}-SiO$_2$ surface:
We observe that in the first layers, due to strong molecule-substrate interactions the molecules first grow in chiral (fan-like) structures, where each consecutive molecule has a higher angle, supported by molecules lying underneath.
Subsequently deposited molecules bind to the already standing molecules of the chiral structures until all molecules are standing.
The growth of chiral islands is the main mechanism for growth of the fluorinated \textit{p}-6P derivative, while the \textit{p}-6P, due to the lower interaction with the underlying substrate, forms less chiral structures.
This leads to a lower energy barrier for step-edge crossing for the fluorinated molecules.
We find that partial fluorination of the \textit{p}-6P molecule can in this way significantly alter its growth behaviour by modifying the rough, 3D growth into a smooth, layer-by-layer growth.
This has implications for the rational design of molecules and their functionalized forms which could be tailored for a desired growth behavior and structure formation.
\end{abstract}


\maketitle

\section{Introduction}
Hybrid structures of organic and inorganic semiconductors (HIOS) have shown enormous application potential in recent years~\cite{1367-2630-10-6-065010, 0953-8984-20-18-184008, 0953-8984-22-8-084024}. Combining the favorable properties of individual materials into a single conjugate makes it possible to realize device properties that cannot be achieved with either material class alone. 
In order to control the properties and functions of HIOS, it is necessary to know the molecular structure and to understand the molecule-molecule interactions and the molecule-surface interactions at the hybrid interface during the interface formation.
Then, controlling the properties and functions of HIOS can be accomplished by a combination of chemical functionalization of the organic adsorbates and a careful selection of the underlying surface.

For instance,the hydrogens in the \textit{meta} positions of both terminal phenyl groups of the prototypical organic para-Sexiphenyl (\textit{p}-6P) molecule can be substituted by fluor atoms to introduce two local dipole moments at both terminal groups~\cite{doi:10.1021/acs.jpcc.8b03398, miletic_polar}. 
This, in turn, changes the degree of diffusion-anisotropy of the molecule on the inorganic zinc oxide $\left(10\overline{1}0\right)$ surface which leads to differences in the growth morphology~\cite{doi:10.1021/jp507776h}. 
Films of \textit{p}-6P are characterized by a three-dimensional morphology with mound-like crystalline islands and a rather rough surface. 
In contrast, the fluorinated derivative (\textit{p}-6P4F) grows in a layer-by-layer mode  with a smooth, two-dimensional morphology~\cite{C4CP04048A}.
Strong molecule-surface interactions can lead to further interesting growth phenomena: surface induced polymorphs (SIPs)~\cite{doi:10.1002/cphc.200900084, https://doi.org/10.1002/adfm.201503169, https://doi.org/10.1002/smll.201403006, C0JM01577F}, wetting layers, i.e. flat-lying molecules at the interfaces~\cite{doi:10.1021/acs.jpcc.8b11238, muccioli_growth} or the 
formation of defective islands with tilted or disordered edges~\cite{https://doi.org/10.1002/adfm.201402609}. 

Next to the important influence of the underlying surface, the grown morphologies also depend critically on the fabrication conditions including deposition rates, substrate temperature, chamber pressure and thermal treatments.~\cite{doi:10.1021/cm049563q, simulation_deposition_ptcdi, Gerlach, MeyerzuHeringdorf2001, PhysRevB.69.165201, PhysRevLett.96.125504, PhysRevB.67.125406}
For instance, recent studies have reported that by lowering the substrate temperature, adsorbed molecules can be controlled to adopt a lying orientation on an {\it a}-SiO$_2$ surface rather than a standing orientation~\cite{Shioya2019, doi:10.1021/acs.jpclett.9b00304}. 
Nagai \textit{et al.}~\cite{reorientation_barrier} also demonstrated that the ratio of standing molecules to lying molecules increases with increasing temperature, as observed in grown films of pentacene on {\it a}-SiO$_2$ studied by \textit{p}-polarized multiple-angle incidence resolution spectrometry (pMAIRS). This suggests that the nucleation of standing-oriented islands is thermally activated at a constant deposition rate. 
Furthermore, at low enough temperatures, mainly the lying orientation occurs and the final film structure consists of lying molecules. 
Nagai \textit{et al.} further observed that the temperature dependence of the probability to nucleate in standing orientation has an Arrhenius behaviour, from which the collective energy barrier for reorientation can be deduced. 
Interestingly, they observed increased migration of adsorbates to distant existing nuclei at higher temperatures enabled by the increased surface diffusion.
Under diffusive conditions at sufficiently high temperatures, molecules can form larger, stable islands with optimal configuration which facilitates the orientational change.
The temperature-dependent evolution of the island size can be explained by the diffusion-mediated growth model, as reported by Tejima \textit{et al.}~\cite{Tejima}. 
In this model, the nucleation density depends on the diffusivity of molecules and the deposition rate. When the deposition rate is constant, the island size increases with the growth temperature because island growth is promoted by high diffusivity.

Although there is an ample amount of experimental studies on growth of organic molecules with different structural and chemical properties, insights on the microscopic features and nanoscale dynamics of molecular growth and thin film formation of HIOS are still rare.
Classical molecular dynamics computer simulations can circumvent experimental limitations and provide an exact picture of the processes involved on the molecular scale.
The computational cost to approach experimental time and length scales however is high and the employed force fields must be good enough to reproduce spontaneously self-assembled room-temperature solid crystals on the inorganic surfaces, like they exist in experimental reality. 
This requires an adequate sampling of the vast configurational space related to the formation of complex molecular structures.
We have previously proposed a force field that is capable of providing a spontaneously self-assembled room-temperature solid \textit{p}-6P crystal with unit-cell parameters in agreement with experimental measurements~\cite{CGD}. 
When it comes to the molecular self-assembly on inorganic surfaces, there is a requirement on the force field to reproduce the correct organic-organic as well as inorganic-organic interactions. 
Another challenge for the current simulation methods is the limited simulation time, which may not be sufficiently long in order to allow the strongly attractive molecules to arrange into ordered positions on the surface. 
Recent simulation studies indicate that relevant growth events can be successfully reproduced on time scales accessible in atomistic simulations, at least for pentacene and sexithiophene molecules~\cite{doi:10.1002/adma.201101652,pizzirusso}. 
Muccioli \textit{et al.}~\cite{doi:10.1002/adma.201101652} demonstrated that in pentacene deposition on a C$_{60}$ crystal surface, the molecules self-assemble into crystal nuclei, closely resembling the bulk crystal structure. The authors stress that the observed deviations in the self-assembled crystal structures could originate from surface distortions or imbalances in the chosen force field. 
Roscioni \textit{et al.}~\cite{muccioli_growth} additionally showed that the vapor-deposition of pentacene molecules onto the amorphous silica surface results in the correct room-temperature bulk crystal structure in the first two adsorbed layers, with unit-cell dimensions consistent with experimental measurements.
Furthermore, Pizzirusso \textit{et al.}~\cite{pizzirusso} showed that sexithiophene molecules spontaneously arrange into an ordered crystal-like structure at room-temperature, consistent with experimental densities and global molecular orientations. 
Simulations of sexithiophene on C$_{60}$ by D'Avino \textit{et al.}~\cite{https://doi.org/10.1002/adfm.201402609} revealed the spontaneous formation of wetting layers and of chiral, propeller-like distorted, crystalline islands with tilted edges. When enough molecules are deposited, the chiral phases expand into crystalline monolayers with well-defined and uniform molecular orientations.

Despite these efforts, important questions still remain. Why do certain chemical modifications such as the fluorination of the \textit{p}-6P alter the growth modes and the structures of the thin films? What mechanisms lead to the change from a 3D growth morphology to a 2D morphology?  What role do the molecule-surface interactions play therein, compared to the molecule-molecule interactions? 
In this paper we employ all-atom molecular dynamics simulations to investigate the influence of change in polarity on nucleation and growth of the \textit{p}-6P and its symmetrically fluorinated derivative \textit{p}-6P4F on the amorphous {\it a}-SiO$_2$ surface. We simulate the growth of up to three complete layers and characterize the layer structures and the molecule-surface interactions in order to detect the driving mechanisms that govern the nucleation and growth of these molecules.

First, we simulate the growth of the first monolayer and analyze the average height, surface roughness and average layer inclination as a function of the surface density of the deposited molecules. We continue the study by depositing molecules onto the first layer and observe and analyze the formation of the second and third molecular layer.  
Then, we provide insight into the surface diffusion dynamics and calculate the binding energies of the molecule to the underlying layers. Finally, we measure the unit-cell parameters of the grown crystalline layers. 
We find that before each layer is completed, the first molecular clusters form chiral structures, if the molecule-surface interactions are strong enough compared to the molecule-molecule interactions. 
The chirality facilitates not only the eventual rise of the molecules but also the diffusion over the molecular terraces.
Once the molecule-surface interactions become too weak, as it happens with the \textit{p}-6P in the second molecular layer but not with the \textit{p}-6P4F, the chirality vanishes long before the layer is completed.
Without the chirality, the energy barrier for crossing the mound terraces becomes prohibitive. 
We finally propose this mechanism as the reason for the three dimensional mound growth in the case of \textit{p}-6P in contrast to the smooth layer-by-layer growth of \textit{p}-6P4F.

\section{\label{sec:methods}Methods} %%%%%%%%%%%%%%%%%%%%%%%%%%%%%%%%%%%%%%%%%%%%%%%%%%%%%%%%%%%%%%%%%%%%%%%%%%%%%%%%%%%%%%%%%%%%%%%%
\subsection{\label{sec:sim_det}Molecular models and simulation details}
\begin{figure*}[ht!]
\includegraphics[width=1\textwidth]{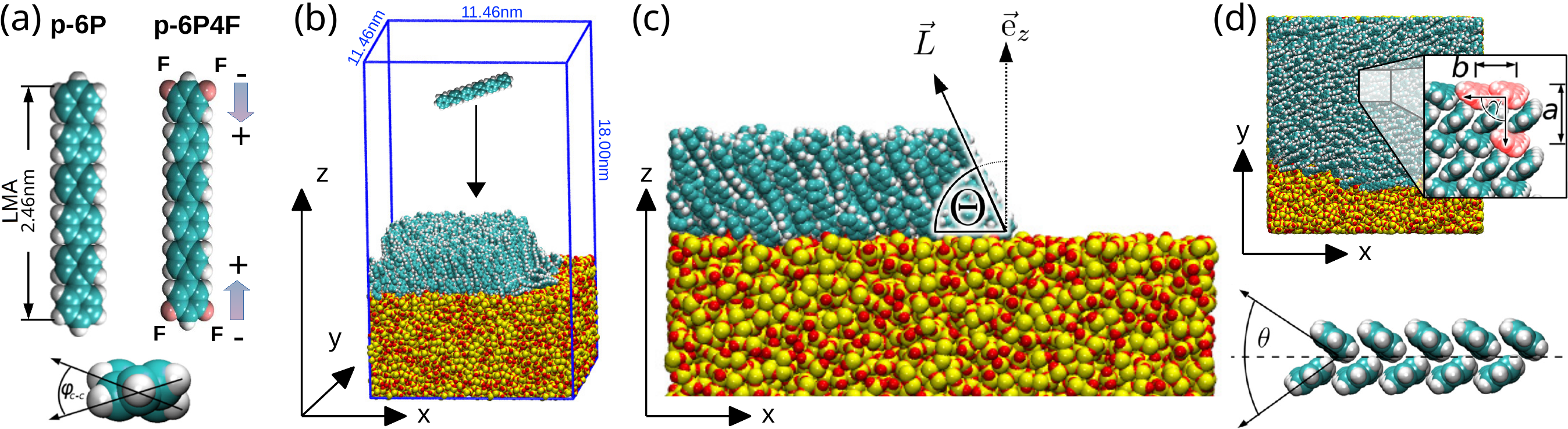}
\caption{Illustrations of the simulation setup, the molecules of interest and
important structural and crystallographic parameters.
a) The conjugated organic para-sexiphenyl molecule (\textit{p}-6P) and
its fluorinated derivative \textit{p}-6P4F. The long molecular axis
(LMA) is defined as the axis which runs through both terminal carbons of
the molecule. The fluorination introduces local dipole moments at the
two head groups of the  \textit{p}-6P4F molecule. For a detailed
distinction between the two molecules see
reference~\cite{miletic_polar}. The front view illustrates the
torsional angle $\varphi_{C-C}$ between neighboring phenyl rings.
b) The simulation setup for the deposition simulations.
c) Projection of the simulation box onto the $x-z$ axis. The angle
$\Theta$, between the vector $\vec{L}$ and the inorganic surface plane,
describes the inclination of the molecules in the grown molecular layer.
The vector $\vec{L}$ is a vector parallel to the LMA of the molecules.
The surface plane is defined by the unit vector in the direction of the
z-axis $\vec{{\rm e}}_{z}$.
d) Projection of the simulation box onto the $x-y$ axis. The inset shows
a schematic illustration of  the lattice parameters $a$, $b$ and the
monoclinic angle $\gamma$ between the crystallographic  $a$- and
$b$-axis. The grown molecular layers possess a herringbone structure
with the herringbone-angle $\theta$ between the molecular planes of the
two molecules of each base.}
\label{fig:illustrations}
\end{figure*}
The simulation setup is illustrated in Fig.~\ref{fig:illustrations}.
The atomic configuration of a silica slab was exported from the VMD molecular visualisation software using the Inorganic Builder tool \cite{HUMP96}. 
The slab with dimensions $11.46 \times 11.46 \times 5.67$ nm$^{3}$  
was equilibrated for 5 ns, with a time step of  0.002 ps, in contact with a heat bath at 300 K. 
The simulation involved 15488 silicon and 30976 oxygen atoms maintained at the bulk density of 2.18 g/cm$^{3}$, in agreement with the experimental density of 2.20 g/cm$^{3}$ \cite{WRIGHT199484}. 
Corresponding radial distribution functions for Si-Si, O-O and Si-O atom pairs are in agreement with X-ray diffraction experiments \cite{WRIGHT199484} and molecular dynamics simulation results from \cite{doi:10.1063/1.463056} (provided in the Supporting Information). 
The interatomic interactions within the amorphous {\it a}-SiO$_2$ are modelled by Lennard-Jones (LJ) and Coulomb interactions to account for van der Waals and electrostatic interactions, respectively.
The partial charges placed on the individual Si and O atoms together with the LJ potential parameters are taken from \cite{doi:10.1021/jp109446d}. 
For modelling the intramolecular and LJ interactions of the \textit{p}-6P and \textit{p}-6P4F molecules, the generalized Amber force field (GAFF) \cite{JCC:JCC20035} is employed. 
The atomic partial charges were calculated with the Gaussian 09 Software using the B3LYP functional and the cc-pVTZ basis set with the electrostatic potential fit (ESP) method.
The LJ interactions between the molecules and the {\it a}-SiO$_2$ surface are governed by the Lorentz-Berthelot combination rules.

To ensure an adequate sampling of the phase space, the time evolution of the position $\vec{r_{i}}$ of each atom $i$ is described by the Langevin equation of motion
\begin{equation}
m_i\frac{d^{2}\vec{r_{i}}}{dt^{2}} = -m_i\xi_i\frac{d\vec{r_{i}}}{dt}+\vec{F_{i}}+\vec{R_{i}}  ,  
\label{eq1}\end{equation}
where $m_i$ is the atomic mass, $\xi_i$ is the friction constant in units of ps$^{-1}$, $\vec{F_{i}}$ is the force acting on atom $i$ due to all other atoms, and $\vec{R_{i}}(t)$  is the random force obeying the fluctuation-dissipation theorem~\cite{doi:10.1021/ct700301q}.
The equation of motion is integrated using a leapfrog algorithm with a time step of 2 fs. 
The long range Coulomb interactions in the system are computed by the Particle Mesh Ewald (PME) method~\cite{doi:10.1021/ct700301q}, using a Coulomb cutoff distance of 1 nm with interpolation order 4 and 30 $\times$ 20 $\times$ 35 grid points in $x$-, $y$- and $z$-directions. 
For the LJ interactions a cutoff of 1.3 nm was applied.

Atoms in the bottom layer of the surface of thickness 1 nm were simulated fixed to their initial relaxed positions, to serve as a structural template for the thermalized amorphous {\it a}-SiO$_2$ surface.
The simulation box was enlarged in the $z$-direction, creating  an  empty  region  where  \textit{p}-6P could  be  deposited, with a final box size of $11.46 \times 11.46 \times 18.00$~nm$^{3}$ and periodic boundary conditions only applied in the $x$- and $y$-directions.   

\subsection{\label{sec:deposition_procedure}Simulation of the experimental deposition process}
The experimental deposition process is simulated as follows.
After an equilibration simulation of the bare {\it a}-SiO$_2$ surface, a single molecule is inserted into the top of the simulation box. The initial orientation and position of the molecule is random albeit restricted to the top of the box such that the distance between the newly inserted molecule's atoms and and the rest of the system is at least 5 nm.
At the start of the simulation, a force is exerted on the molecule that causes it to accelerate in the negative $z$-direction (towards the surface, see Fig.~\ref{fig:illustrations} b).
After a certain time $\tau_{{\rm in}}$, the accelerating force is removed from the molecule and a new molecule is inserted, again with a random yet restricted orientation and position and with a downwards acceleration. This process is repeated until three full molecular layers have formed.

The time $\tau_{{\rm in}}$ between two consecutive insertions is restricted due to the computational cost associated with the necessary length scales.
If possible, e.g. during the growth of the first molecular layer, we set $\tau_{{\rm in}}=3000$~ps (corresponding to the deposition rate $\tau_{{\rm in}}^{-1}$).
When more molecules are involved, we use $\tau_{{\rm in}}=300$~ps, which corresponds to a higher deposition rate. We also investigated what happens if $\tau_{{\rm in}}=30$~ps.

The small acceleration that every newly inserted molecule is exposed to, is necessary to slowly bring the molecule within range of the molecule-surface interactions.
If $\tau_{{\rm in}}=3000$~ps, we have enough time to accelerate relatively slowly ($a_{z}=-0.01$~nm/ps${^2}$). 
In the simulations with $\tau_{{\rm in}}=300$~ps and $\tau_{{\rm in}}=30$~ps, the acceleration has to be increased to $a_{z}=-0.03$~nm/ps${^2}$ and $a_{z}=-0.05$~nm/ps${^2}$, respectively.

The deposition simulation is performed at a temperature of 575~K. This temperature is higher than those usually employed in deposition experiments~\cite{ATHOUEL199635}. 
However, it has been shown that at this temperature the herringbone structure of the \textit{p}-6P is still preserved~\cite{doi:10.1021/cg500234r}.
We adopt such a high temperature in our simulations to speed up the molecular motion, in order to prevent molecules from getting kinetically trapped due to the high (compared to experiments) deposition rate~\cite{clancy_diff}.
If the deposition simulation is directly performed at $T$=300~K, the molecules self-assemble into irregular structures and layer-by-layer growth is no longer observed. 
Yet, by gradual temperature annealing from $T$=575~K to $T$=300~K we can equilibrate the final structures into the correct room temperature crystal structure.

\subsection{\label{sec:Binding_sio}Single molecule surface binding}

To interpret the structures of the grown molecular layers, it is imperative to know the free energy for the binding/unbinding process of a molecule to its substrate. 
The free energy for the binding/unbinding process, $\Delta F_{b}$,  is estimated from the potentials of mean force (PMF) as $\Delta F_{b}=F_{b}(z_{0})-F_{b}(z_{\rm max})$ where $z_{0}$ and $z_{\rm max}$ are the global minimum of the curve and the reference distance along the $z$ coordinate, respectively. 
The reference distance $z_{\rm max}$ is defined as the distance between molecule and surface where both are so far apart that they do not exert influence on each other.
The PMFs are calculated from steered MD simulations~\cite{gromacs5} at $T$ = 575~K, where the center-of-mass (COM) of the molecule is connected to a virtual site via a harmonic potential with the spring constant $k = 5000$~kJ~mol$^{-1}$~nm$^{-1}$, while the virtual site moves away from the surface with a constant velocity of $10^{-4}$~nm~ps$^{-1}$. 
The chosen value for the pulling velocity is lower by at least a factor of 5 compared to previously employed pulling velocities in studies of binding of structuraly similar or more complex ligands to protein structures~\cite{doi:10.1021/ci4003574,doi:10.1021/ja100259r}. 
It was found to be a good compromise between the accuracy and the simulation speed.
The PMF is then obtained from the integral of the net force acting on the harmonic spring while the molecule is pulled away from the surface.  
The origin $z=0$ is defined as the $z$-coordinate of the highest atom of the silica surface.

\subsection{\label{sec:Diff}Surface diffusion} %%%%%%%%%%%%%%%%%%%%%%%%%%%%%%%%%%%%%%%%%%%%%%%%%%%%%%%%%%%%%%%%%%%%%%%%%%%%%%%%%%%%%%%%%%%%%%%%
As molecules are deposited one by one on the surface, each molecule diffuses on the surface before it eventually integrates into the nearest molecular cluster.
Thus, understanding the subtle differences in diffusion between the \textit{p}-6P and \textit{p}-6P4F can help us to understand the differences in the growth modes between \textit{p}-6P and \textit{p}-6P4F.
The total long-time in-plane diffusion coefficients, $D^{\rm tot}$, averaged over the $x$- and $y$-direction of the molecular motion, are obtained from the mean squared displacements (MSDs) of the molecular COM from the simulated trajectories, via
\begin{equation}
\left\langle \left[x(t)-x(t_{0})\right]^{2}+\left[y(t)-y(t_{0})\right]^{2}\right\rangle = \lim_{t \to \infty} 4 D^{\rm tot} t.
\label{MSD}
\end{equation}
where $x(t)$ and $y(t)$ is the coordinates of the molecular COM at time $t$ and $\left\langle \right\rangle$ denotes the ensemble average.

\section{Results}

\subsection{Formation of the first monolayer}
\label{1ML-form}
\begin{figure*}[ht!]
\includegraphics[width=16.7cm]{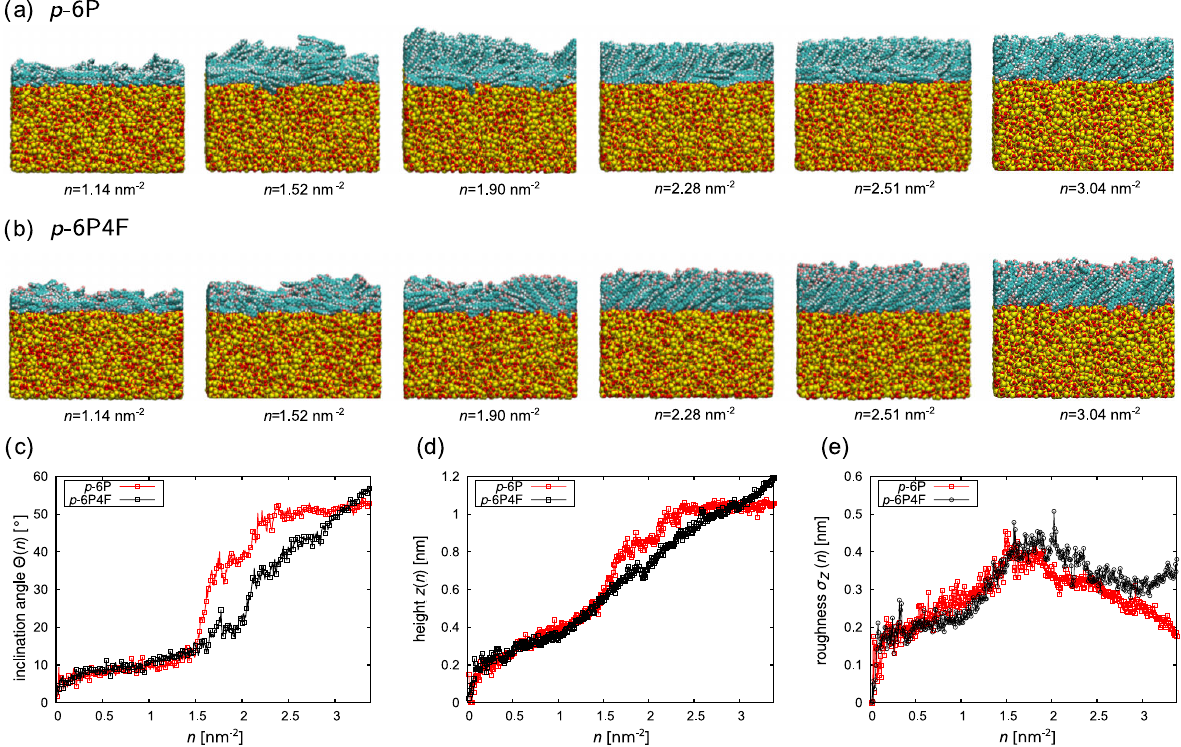}
\caption{a) Simulation snapshots of \textit{p}-6P molecules deposited onto the silica surface, at $T$ = 575 K with a deposition rate of~3000~$^{-1}$ ps$^{-1}$. 
b) Simulation snapshots of \textit{p}-6P4F molecules deposited onto the silica surface, at $T$ = 575 K with a deposition rate of~3000~$^{-1}$ ps$^{-1}$. 
c) The inclination angle $\Theta(n)$ of the 1ML as a function of the surface density $n$ of the deposited \textit{p}-6P and \textit{p}-6P4F molecules.
d) The height $z(n)$ of the 1ML.
e) The roughness $\sigma_{z}(n)$ of the 1ML.}
\label{3000ps}
\end{figure*}
In this section we examine the influence of change in molecular polarity on the formation and structural properties of the first monolayer (1ML).
The deposition simulation is performed by inserting one molecule into the simulation box every 3000 ps (that corresponds to a rate of 3000$^{-1}$ ps$^{-1}$).

In Fig.~\ref{3000ps} a)-b) we show simulation snapshots for the \textit{p}-6P and \textit{p}-6P4F molecules.
Fig.~\ref{3000ps} c) shows the average inclination angle $\Theta(n)=\frac{1}{N}\sum_{i}^N\Theta_{i}(n)$ as a function of the surface density $n$, which is defined as the number of already deposited molecules $N$ divided by the area of the surface ($11.46 \times 11.46$~nm$^2$).
The inclination angle $\Theta_{i}(n)$ of a single molecule $i$ is defined as the angle between the LMA and the surface plane via
\begin{equation}
\Theta_{i}(n)=\arcsin\left(\frac{\vec{L}_i(n)\cdot\vec{{\rm e}}_{z}}{\left|\vec{L}_i(n)\right|}\right),
\end{equation}
where $\vec{L}_i(n)$ is a vector parallel to the LMA, pointing from one terminal carbon of the molecule $i$ to the other. The surface normal is described by $\vec{{\rm e}}_{z}$, which is the unit vector in the direction of the $z$-axis.
Fig.~\ref{3000ps} d) shows the average height $z(n)$ of the 1ML as a function of surface density $n$. 
The height of the deposited layer(s) is defined as $z(n)=1/N\sum_{i}^{N}z_{i}(n)$, where the height $z_{i}(n)=z_{i}^{{\rm COM}}(n)-z^{{\rm SiO_{2}}}$ of a molecule $i$ is the difference between the $z$-coordinate of the molecule's COM and the $z$-coordinate of the highest atom of the silica surface. 

Finally, Fig.~\ref{3000ps} e) shows the roughness $\sigma_{z}(n)$ of the monolayer, calculated as the standard deviation of the 1ML height
\begin{equation}
\sigma_{z}(n)=\sqrt{\frac{1}{N-1}\sum_{i}^{N}\left(z_{i}(n)-z(n)\right)^{2}}.
\end{equation}

Up until the first 150 molecules are deposited (corresponding to a surface area density of $n=1.14~$nm$^{-2}$), all molecules are lying flat on the surface. 
At this stage, the molecules resemble a molecular liquid and homogeneously wet the surface. 
However, as molecules slide over each other, they start to create chiral, fan-like, structures, where each consecutive molecule has a higher angle because it is propped by all the other molecules lying underneath. 
This can be nicely observed in the $n=1.52~$nm$^{-2}$ snapshot for the case of \textit{p}-6P or in the $n=1.90~$nm$^{-2}$ snapshot in the case of \textit{p}-6P4F, for example. 
The more molecules are integrated into these chiral fans, the more molecules will eventually be standing instead of lying, and the average inclination angle of the 1ML increases (see Fig.~\ref{3000ps} c). 
This process goes hand in hand with the unbinding of the molecules from the surface while they instead bind to the already standing molecules.
After about 400 molecules are deposited on the surface (corresponding to a surface area density of $n=3.04~$nm$^{-2}$), almost all the molecules of the 1ML are standing and the 1ML is complete.

When the average height reaches $z\approx1.2$ nm, the majority of the molecules are in the upright standing configuration on the surface (see Fig.~\ref{3000ps} d).
The roughness of the 1ML increases as molecules are deposited on the surface and as the amount of chiral structures increases (see Fig.~\ref{3000ps} e). 
The roughness reaches its peak when the orientational change propagates through the 1ML and decreases after about 350 molecules ($n=2.66~$nm$^{-2}$) are deposited and the chiral structures are replaced by regular crystallinity, in both cases.

\begin{figure*}[ht]
\includegraphics[width=17.0cm]{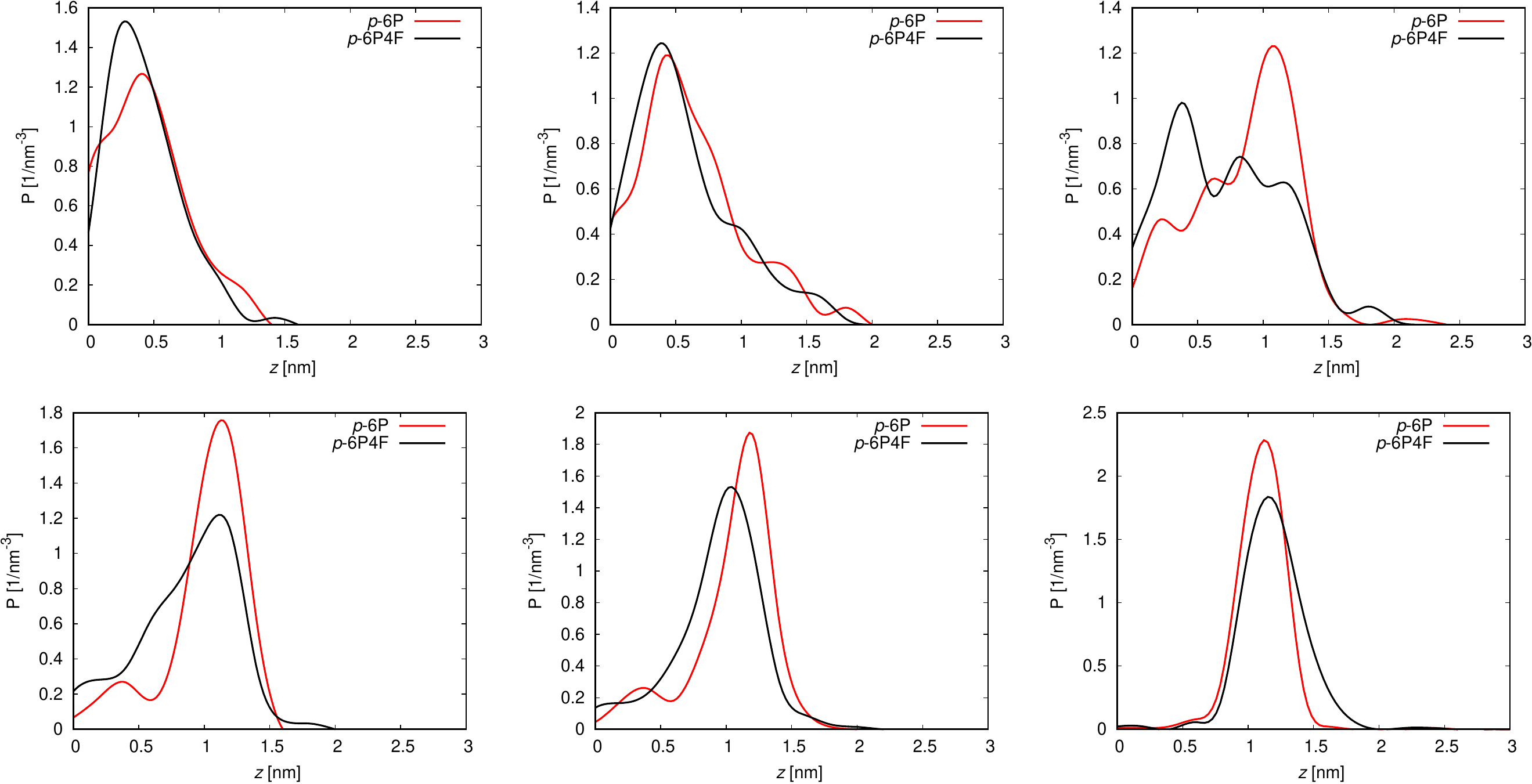}
\caption{Probability distributions of the molecular COM as a function of the layer
 height for the \textit{p}-6P and \textit{p}-6P4F molecules deposited onto
the silica surface with a deposition rate of
3000$^{-1}$ ps$^{-1}$. The plots correspond to the surface density of $n$=1.14, 1.52, 1.90, 2.28, 2.51 and 3.34 nm$^{-2}$, with $n$ increasing from left to right. The average COM $z$-coordinate of 1.22 nm corresponds to half of the
molecular length (corresponding to the monolayer height with molecules in the upright standing configuration).
}
\label{prob_dens}
\end{figure*}
Fig.~\ref{prob_dens} shows probability distributions of the molecular COM as a function of the layer height for the \textit{p}-6P and \textit{p}-6P4F molecules deposited onto the silica surface with a deposition rate of 3000$^{-1}$ ps$^{-1}$. 
The plots correspond to the surface densities of $n$=1.14, 1.52, 1.90, 2.28, 2.51 and 3.34 nm$^{-2}$, with $n$ increasing from left to right. 
The average COM $z$-coordinate of 1.22 nm corresponds to half of the molecular length (corresponding to the monolayer height with molecules in the upright standing configuration).
The peak in the red curve (\textit{p}-6P) progresses faster along the $z$-axis compared to the black curve, indicating that the \textit{p}-6P changes the orientation sooner compared to the other molecule. 
This result, based on a single deposition simulation, could indicate that the \textit{p}-6P requires a smaller critical nucleus size to have an orientational change. However, additional deposition simulations are required to support this conclusion.
The second molecular layer (2ML) then starts forming after the 1ML is completed.

\subsection{Formation of the second monolayer}
\label{2ML-form}
\begin{figure*}[ht]
\includegraphics[width=16.6cm]{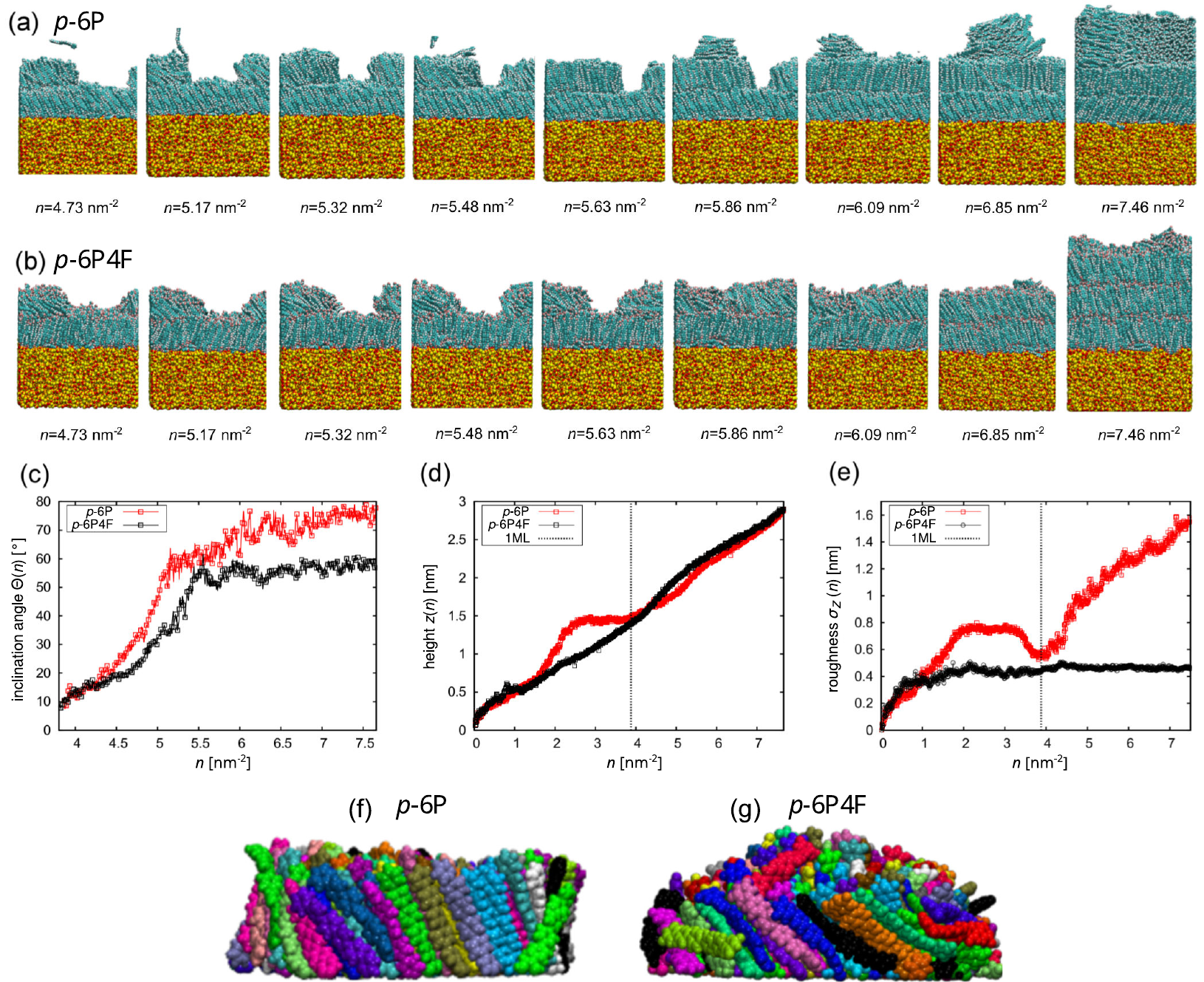}
\caption{a) Simulation snapshots of \textit{p}-6P molecules deposited onto the silica surface, at $T$ = 575 K with a deposition rate of 300$^{-1}$ ps$^{-1}$. 
The number of molecules increases from 600 to 1500 from left to right.
b) Simulation snapshots of \textit{p}-6P4F molecules deposited onto the silica surface, at $T$ = 575 K with a deposition rate of 300$^{-1}$ ps$^{-1}$. As in the case of the \textit{p}-6P, the number of molecules increases from 600 to 1500 from left to right.
c) The inclination angle $\Theta(n)$ of the deposited layers as a function of the surface density $n$ of the deposited \textit{p}-6P and \textit{p}-6P4F molecules.
d) The height $z(n)$ of the deposited layers.
e) The roughness $\sigma_{z}(n)$ of the deposited layers.
f-g) Side view of the chiral crystalline islands formed by the \textit{p}-6P and \textit{p}-6P4F molecules adsorbed on the respective 1MLs, with surface densities of $n$=5.10 nm$^{-2}$. For clarity the 1ML and silica substrates are not shown and molecules are drawn with different colors.}
\label{2ml}
\end{figure*}

\begin{figure*}[ht]
\includegraphics[width=16.6cm]{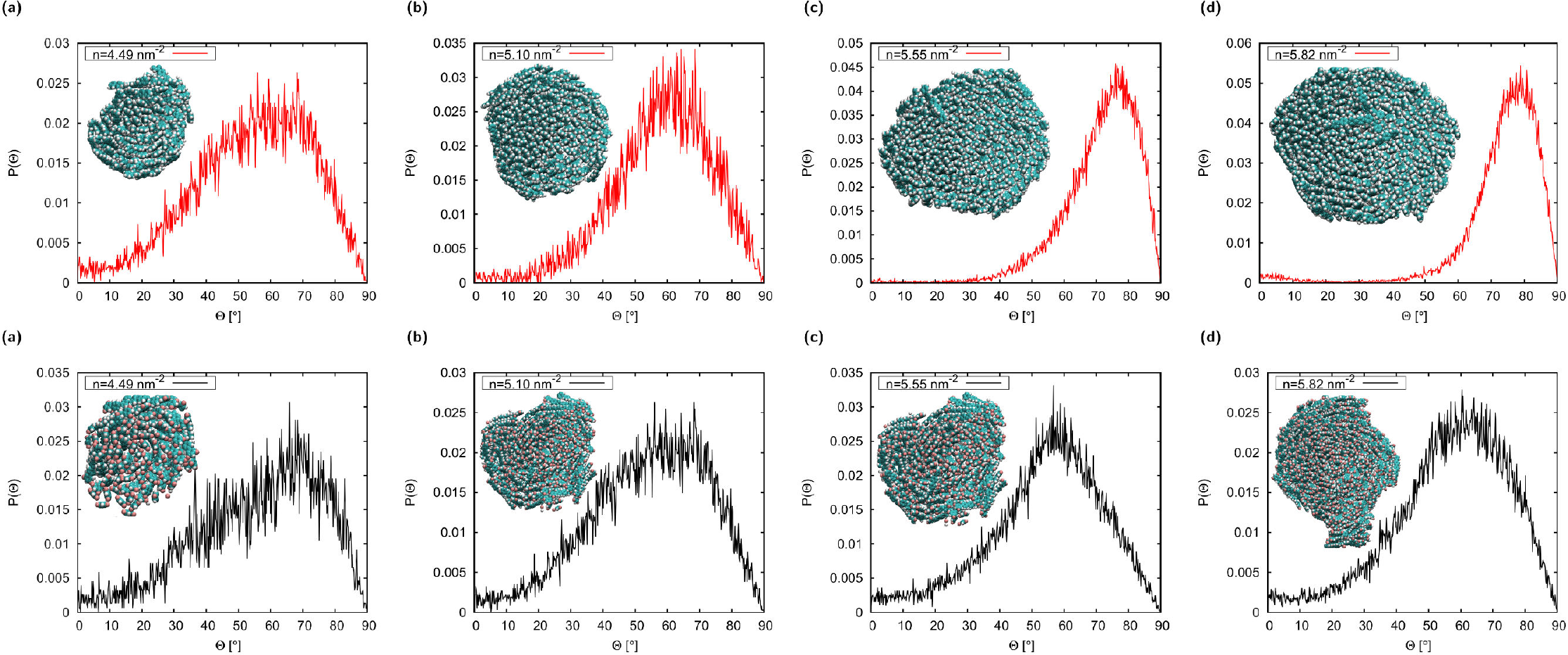}
\caption{Probability distributions of the average inclination angle for the molecules in the 2ML. The insets of the figures are showing representative
snapshots of grown islands observed from the top, displaying the chiral island edges.}
\label{chirality}
\end{figure*}

In this section we examine the influence of change in molecular polarity on the formation and structural properties of two monolayers.
Due to the increase in the computational cost with an increase in the system size, we have to simulate the growth of molecules with a faster rate of 300$^{-1}$~ps$^{-1}$.
The consequences of this choice are discussed in section~\ref{dep_rate}.
The deposition of both \textit{p}-6P4F and \textit{p}-6P leads to the formation of a crystalline 1ML with standing and tilted molecules, in accordance with experiments~\cite{C4CP04048A} and with our simulations of the 1ML at a lower rate in the previous section.
Fig.~\ref{2ml} shows representative snapshots of the second and third monolayers consisting of the a) \textit{p}-6P and b) \textit{p}-6P4F. 

In the second monolayer (2ML), smooth, layer-by-layer growth is observed in the case of \textit{p}-6P4F.
The process is very similar to the growth of the 1ML with the difference that the molecules start to form chiral structures on the 2ML at a much lower surface coverage compared to the 1ML growth on bare silica. 
This is because, in contrast to the growth of the 1ML on silica, the molecules do not wet the entire surface before clustering to the previously described fan-like shapes that eventually lead to the first standing molecules.
Instead, the \textit{p}-6P4F molecules form the fan-like clusters very soon after being deposited (see Fig. \ref{2ml} g)). 
Beyond that, the 2ML grows just like the 1ML did: 
Once the clusters become big enough, the number of upright standing molecules increases until it becomes energetically favorable for the other molecules to stand up as well.
After the 2ML is completed, the 3ML proceeds to grow in the same way as the 2ML.

In the case of the \textit{p}-6P, we observe a different behavior during the growth of the 2ML. 
At first, when the first 50 molecules are deposited, the first clusters grow similarly to \textit{p}-6P4F: 
They form fan-like structures of molecules with different inclination angles, sometimes also resembling wooden logs arranged in a campfire (see Fig. \ref{2ml} f)). 
However, the 2ML loses its chiral property at a much lower coverage than \textit{p}-6P4F, as can be seen in Fig.~\ref{chirality}, which shows probability distributions of the inclination angle of the molecules belonging to the 2ML. 
At a surface coverage of 5.55 nm$^{-1}$, i.e., before the 2ML is complete, most \textit{p}-6P molecules are already standing.
From then on, \textit{p}-6P  grows into irregular structures, which has been previously observed in growth experiments of \textit{p}-6P on the ZnO $\left(10\overline{1}0\right)$ surface~\cite{C4CP04048A}. 
As shown in the 6'th snapshot from the left in Fig.~\ref{2ml} a), the newly deposited molecules that get adsorbed on the top of the 2ML do not diffuse down the step-edge to integrate into the 2ML, but form new clusters which are not chiral anymore. 
They form stacks of equally oriented, lying molecules which only grow bigger in size when more molecules are adsorbed. 
Eventually, the 2ML gets filled with those molecules that are, by chance, inserted into the simulation box right above the gap of the 2ML.
By that time, the cluster on top of the 2ML has grown significantly, causing a strong increase of the roughness (see Fig.~\ref{2ml} d).
Within the simulation time, those clusters do not change their orientation.
Thus, what should essentially be the 3ML, instead becomes a structure of lying molecules, growing vertically. 

Chirality at the edges of the \textit{p}-6P4F islands facilitates downward mass transport and conversely, layer-by-layer growth.
Because of the quite diverse morphology of the chiral terrace step-edge and the relatively small inclination angles of the step-edge molecules, the diffusing molecules have a relatively high chance to find an energy pathway low enough for them to descend the step-edge and integrate into the existing layer. 
It has been shown that the barrier for the terrace crossing strongly depends on the average inclination of the island and
decreases with a decrease in the inclination angle~\cite{hlawacek:sience}.
Also, based on the results shown in Fig.~\ref{chirality}, the average inclination of the \textit{p}-6P4F islands is lower compared to the \textit{p}-6P for the same surface density.
However, in the case of the \textit{p}-6P, weaker attraction to the surface enhances both the diffusivity and cluster formation, but reduces the chirality at the island edges.
There is a threshold surface density where this effect becomes especially prominent, which is around $n$=5.82 nm$^{-2}$, with an average island inclination angle of around 76\textdegree. 
Thus, with less inclined islands, where the chirality is significantly reduced at the island edges at the surface densities higher than $n\geq$~5.82 nm$^{-2}$,
\textit{p}-6P experiences higher barrier for the terrace crossing compared the \textit{p}-6P4F, which hampers layer-by-layer growth starting from the 2ML.

This difference in growth mode between \textit{p}-6P and \textit{p}-6P4F can be explained by analyzing
the diffusion coefficient and the binding energies of the adsorbed molecules to the respective surfaces.
In the following sections we will study these properties and find a relationship between the type of surface, the degree of chirality and the growth mode.

\subsection{Interface layers}
\label{IL}
We observe the presence of horizontal molecules between the silica surface and 1ML, that belong to the interface layer (IL). The existence of one or several ILs or wetting layers of flat molecules was reported in several studies
for molecules such as pentacene, anthracene and rubicene~\cite{Meyer_zu_Heringdorf_2008, doi:10.1021/ci4003574, muccioli_growth} 
where in some cases the insertion of organic ILs has been utilized as a technique to passivate or manipulate the interaction with metallic surfaces.
In our case, the number of molecules that belong to the interface layer increases with an increase in the number of polar groups in the molecule
and amounts to 3 and 8 molecules for the \textit{p}-6P and \textit{p}-6P4F in the 1ML, respectively, while the 2ML proceeds to grow without a wetting layer. 
This might be a consequence of the stronger interaction of molecules with silica compared to their respective 1MLs (see Fig.~\ref{binding_summary}).
However, as the wetting layers in this case are characterized each based on a single deposition simulation, our result can provide an inspiration for studying wetting layers from multiple deposition simulations for statistically meaningful results. 

Some molecules belonging to the IL are able to diffuse on the bare surface but become kinetically trapped
below the growing 1ML terrace, while some others are immobile on specific positions on the silica surface. 
As it was previously suggested that these immobile IL molecules could play a role in the formation of the 1ML on silica~\cite{muccioli_growth}, it is crucial to study how the single molecule surface diffusion
is related to the formation of ILs. 

\subsection{Single molecule surface diffusion}
\label{diffusion}
\begin{table} [ht]
\caption{Diffusion coefficients $D^{\rm tot} (T)$ in units of nm$^2$ps$^{-1}$ calculated according to eq.~\ref{MSD} for single \textit{p}-6P and \textit{p}-6P4F molecules diffusing on the {\it a}-SiO$_2$ and 1ML at $T=575$~K.}
{\footnotesize %
\centering
\setlength{\tabcolsep}{0.125cm}
{\renewcommand{\arraystretch}{1.5}
\begin{tabular}{cccccccc}
\hline
\hline
  &     on {\it a}-SiO$_2$ &  &  & on 1ML\tabularnewline 
\hline 
  T (K)   &  \textit{p}-6P                    & \textit{p}-6P4F              &   &  \textit{p}-6P   & \textit{p}-6P4F\tabularnewline   
\hline 
575 &  1.19$\cdot10^{-4}$     &     1.11$\cdot10^{-4}$   &     &   6.13$\cdot10^{-3}$    &   3.91$\cdot10^{-3}$  \tabularnewline

\hline
\hline
\label{diffusion}
\end{tabular}
}}
\end{table}
When in equilibrium, single molecules are observed to adsorb and diffuse in a flat-lying geometry on the surface. 
To study the surface diffusion, we simulate, at various temperatures, 
(i) a single \textit{p}-6P molecule on the bare silica surface, (ii) a single \textit{p}-6P4F molecule on the bare silica surface, 
(iii) a single \textit{p}-6P molecule on top of the 1ML made of \textit{p}-6P molecules, and (iv) a single \textit{p}-6P4F molecule on top of the 1ML made of \textit{p}-6P4F molecules.
The diffusion coefficients are calculated according to eq.~\ref{MSD}.
The $T$ dependent diffusion coefficients and resulting diffusion energy barriers are discussed in the Supporting Information.

Here, we present in Table \ref{diffusion} the results for $T=575$~K, which are extrapolated using the Arrhenius equation (see Supporting Information). 
As we can observe, on silica, the \textit{p}-6P4F has a lower diffusion coefficient compared to the \textit{p}-6P.
This result indicates that \textit{p}-6P4F molecules are more likely to become trapped at the interface layer during the deposition. 
On the 1ML, the \textit{p}-6P also diffuses faster than the \textit{p}-6P4F, due to its lower surface free binding energy compared to its fluorinated derivatives.
Finally, we see that the diffusion coefficients on the 1ML are about 1 order of magnitude higher compared to the diffusion coefficients on silica.
This result supports our finding that, once the 1ML is finished, the orientational change during the 2ML growth happens sooner (see Fig.~\ref{2ml}) as molecules diffuse significantly faster than on the bare silica and are able to find their binding partners sooner in the given simulation time. 
The average timescale for the diffusion process to be activated on the surfaces is about 1 ns on average\cite{doi:10.1063/5.0024178}. 
This suggests that one should refer to the system-specific diffusion timescale while opting for a deposition rate in the MD simulations of growth. 
Note that in our simulations the deposition time scale of $\tau_{\rm in}=3000$~ps is three times higher than the average diffusion timescale.

\subsection{Surface binding energy}
\begin{figure}[h!]
\centering
\includegraphics[width=8.5cm]{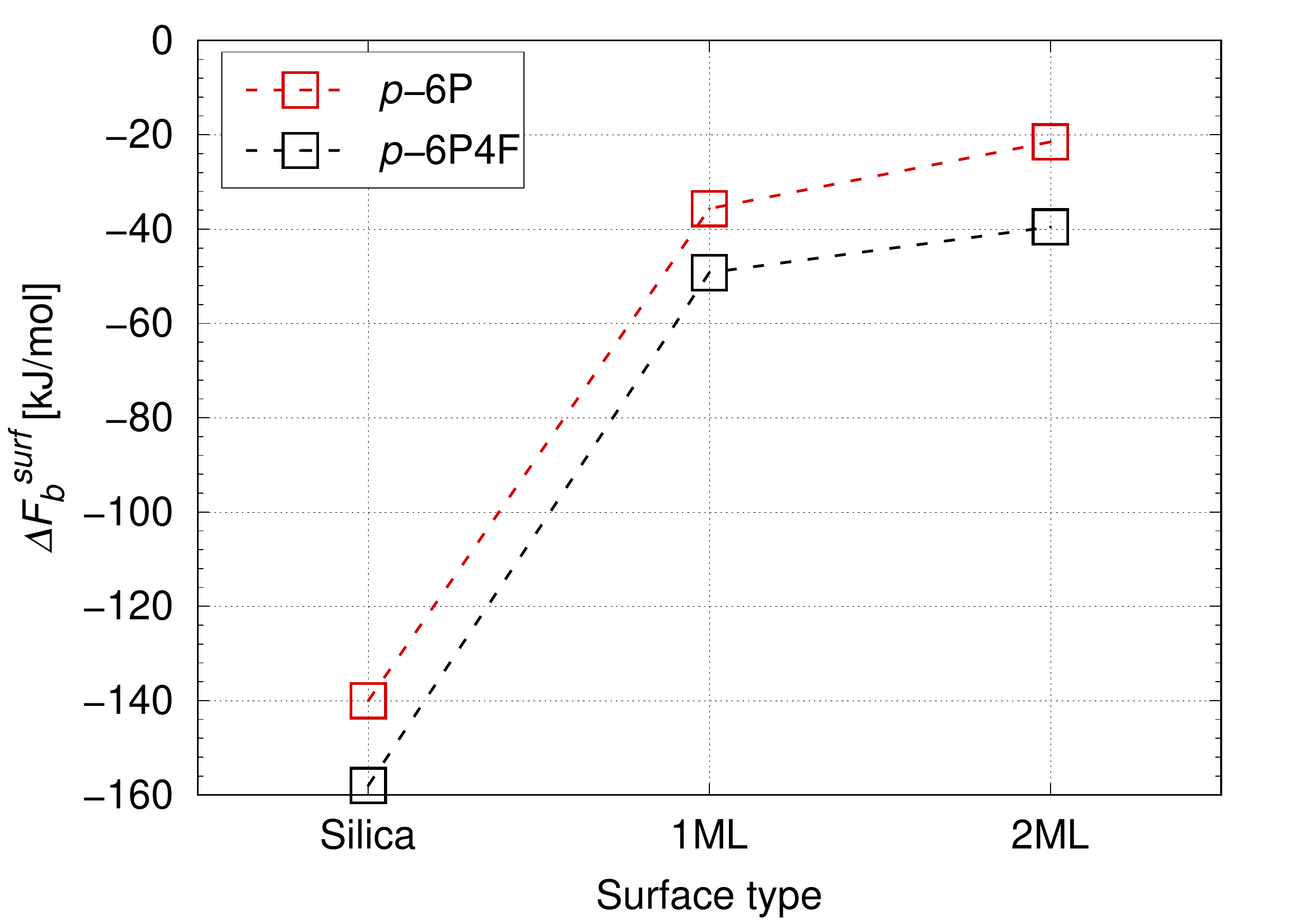}
\caption{Comparison of the binding free energy of the \textit{p}-6P4F and \textit{p}-6P molecules at $T$ = 575 K on three different surfaces: amorphous silica surface, 1ML and 2ML. The \textit{p}-6P4F binds stronger to the silica surface, 1ML and 2ML, compared to the \textit{p}-6P.}
\label{binding_summary}
\end{figure}

Fig.~\ref{binding_summary} compares the binding free energies on three different surfaces onto which molecules are deposited: amorphous silica, 1ML and 2ML.
The binding free energy is defined as a difference between the free energy in the adsorbed state and free energy in the desorbed state (molecule far away from the surface).
We find that the surface binding free energy decreases with each new molecular layer.
Evidently, the \textit{p}-6P4F binds stronger to the silica surface, 1ML and 2ML, compared to the \textit{p}-6P. 
The stronger binding is a consequence of the polar 
head and tail groups in the \textit{p}-6P4F interacting with underlying surface dipoles. As a result, an increase in the number of polar groups also hampers the diffusion on silica 
and also on 1ML and 2ML. This, in turn, increases the number of molecules participating in the wetting layer on the bare silica compared to the MLs.

As an important consequence of the stronger binding to the underlying surfaces, there is a high probability for the \textit{p}-6P4F to grow
into fan-like structures in the early stages of 2ML and 3ML formation. 
Our finding suggests that strong binding to the surface favors chirality, while the chirality facilitates the smooth growth observed during the growth of the 1ML (in case of both molecules) and all subsequent layers of the \textit{p}-6P4F.
As for the \textit{p}-6P, if the binding energy becomes small enough, the molecules lose their chiral state before the layer is fully formed.
This, in turn, increases the inclination of the island and the step-edge barrier, which ultimately hampers the downward diffusion.
On the other hand, chirality at the edges of the \textit{p}-6P4F islands facilitates downward mass transport and
conversely, layer-by-layer growth, as the molecule can always find an energy pathway low enough for it to descend the terrace.

\subsection{Structural properties of adsorbed layers}
\subsubsection{Structural properties of the 1ML}
\label{1ML-str}
\begin{figure*}[ht]
\centering
\includegraphics[width=11.4cm]{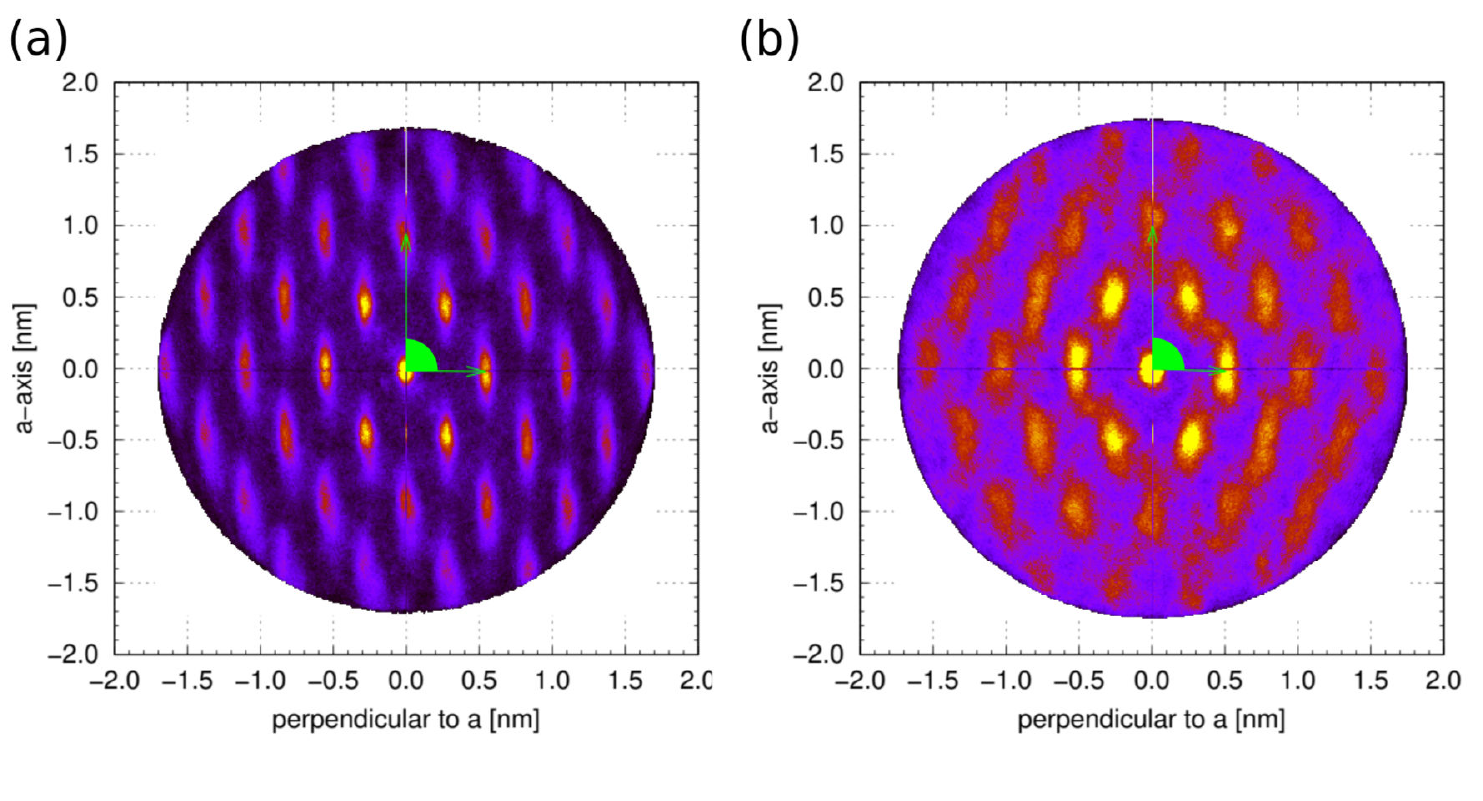}
\caption{Time-averaged in-plane radial density distribution $g(x,y)$ of the molecular COMs, showing the close "environment" of each molecule in the 1ML. a) \textit{p}-6P, b) \textit{p}-6P4F. The arrows show the directions of the unit-cell $a$-axis and the $b$-axis and the angle $\gamma$ between the axes.}
\label{1ML_rdf}
\end{figure*}

After the completion of the 1ML, the final structure is equilibrated with temperature annealing by decreasing the system temperature gradually from $T$=575 K to $T$=300 K.
\begin{table*} [ht]
\caption{Calculated room-temperature (herringbone-phase) properties of the \textit{p}-6P and \textit{p}-6P4F 1ML structures. Obtained results are compared with the results from MD simulations~\cite{doi:10.1021/cg500234r} and experimental~\cite{BAKER19931571} studies of the \textit{p}-6P bulk crystal structure properties at $T$ = 300 K. 
The unit-cell parameters $a$, $b$, $\gamma$, the herringbone angle $\theta$, inclination angle $\Theta$,  and torsion angle $\varphi_{C-C}$ are defined in figure~\ref{fig:illustrations}. $\rho$ is the mass density of the molecules.}
{\footnotesize %
\centering
\setlength{\tabcolsep}{0.125cm}
{\renewcommand{\arraystretch}{1.5}
\begin{tabular}{cccccccccc}
\hline
\hline
                                       & $a$ [nm]           &  $b$ [nm]      &  $\gamma$ [\textdegree] &  $\theta$ [\textdegree] &  $\Theta$ [\textdegree] & $\varphi_{C-C}$ [\textdegree]  & $\rho$ [g/cm$^3$] \tabularnewline
\hline 
\textit{p}-6P                             &   0.95$\pm$0.17    & 0.54$\pm$0.04   & 90.08$\pm$13.30 &  64.30$\pm$19.40           &   34.60$\pm$14.00       & 15.40$\pm$11.00             &  1.31$\pm$0.07\tabularnewline

\textit{p}-6P4F                           &   1.00$\pm$0.06  & 0.50$\pm$0.05   & 91.9$\pm$16.50    &   75.70$\pm$25.10          &   34.50$\pm$13.50       & 18.30$\pm$11.50	     &  1.31$\pm$0.07\tabularnewline 

Ref. \cite{doi:10.1021/cg500234r}     & 0.827$\pm$0.01    & 0.548$\pm$0.01 &   89.80$\pm$3.30    & 61.7$\pm$13.70             &  17.70$\pm$6.00          & 15.70$\pm$7.90	     &  1.29$\pm$0.02\tabularnewline  

Ref. \cite{BAKER19931571}              & 0.809            & 0.557           &    90.00            & 66.00                     &   18.00                    & 20.00 		     &  1.30\tabularnewline 

\hline
\hline
\label{unitcell1}
\end{tabular}
}}
\end{table*}
Table \ref{unitcell1} is showing the room-temperature (herringbone-phase) properties of the \textit{p}-6P and \textit{p}-6P4F 1ML structures. 
Fig. \ref{1ML_rdf} shows a time-averaged in-plane radial density distribution of the molecular COMs
\begin{equation}
g\left(x,y\right)=\frac{1}{n}\left\langle \sum_{i}\delta\left(x-x_{i}\right)\delta\left(y-y_{i}\right)\right\rangle, 
\end{equation}
where $n$ is the number of the deposited molecules divided by the area of the surface and $\left\langle \right\rangle $ denotes the ensemble average. 
To obtain more statistical data, we not only calculate distances between molecules from the same simulation timeframe but also between molecules from different timeframes.

The agreement between MD simulations and experiments in Table~\ref{unitcell1} validates the accuracy of the used force field. 
Even though the growth of organic structures is governed by processes occurring on time and length scales largely exceeding those accessible in our simulations, we emphasize that our high-rate and high-temperature deposition simulations can reproduce the layer-by-layer growth of the first MLs, resulting in a realistic molecular packing.

\subsubsection{Structural properties of the 2ML}
\label{2ML-str}
In this section we investigate the structural properties of the 2ML by quantifying unit-cell parameters. Table~\ref{unitcell2} shows the room-temperature (herringbone-phase) properties of the \textit{p}-6P and \textit{p}-6P4F.
After the completion of the 2ML, the final structure is equilibrated with gradual temperature annealing by decreasing the system temperature from $T$=575 K to $T$=300 K.

\begin{figure*}[ht]
\includegraphics[width=17.4cm]{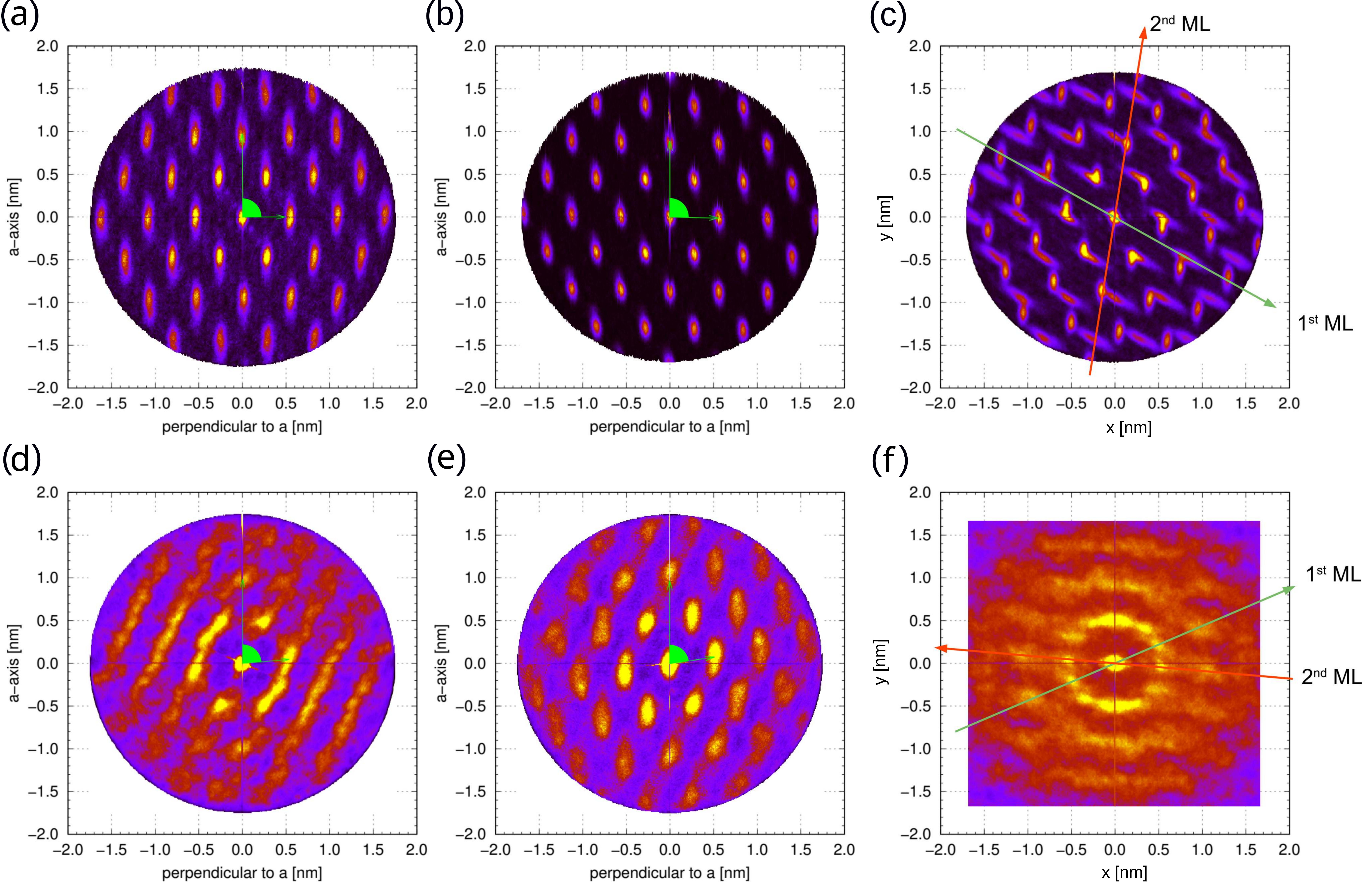}
\caption{Time-averaged in-plane radial density distribution $g(x,y)$ of the molecular COMs, showing the close "environment" of each molecule in 
a) the 1ML of \textit{p}-6P, 
b) the 2ML of \textit{p}-6P,
c) the 1ML and the 2ML of \textit{p}-6P and the relative orientation of the unit-cells between the two layers
d) the 1ML of \textit{p}-6P4F, 
e) the 2ML of \textit{p}-6P4F,
f) the 1ML and the 2ML of \textit{p}-6P4F and the relative orientation of the unit-cells between the two layers.
The arrows in (a),(b),(d),(e) show the directions of the unit-cell $a$-axis and the $b$-axis and the angle $\gamma$ between the axes. The arrows in (c) and (d) show the direction of the unit-cell $a$-axis of each respective layer.}
\label{2ML_dens}
\end{figure*}

\begin{table*} [ht]
\caption{Calculated room-temperature (herringbone-phase) properties of the \textit{p}-6P and \textit{p}-6P4F 2ML structures. Obtained results are compared with the results from the MD simulation~\cite{doi:10.1021/cg500234r} and experimental~\cite{BAKER19931571} studies of the \textit{p}-6P bulk crystal structure properties at $T$ = 300 K.
The unit-cell parameters $a$, $b$, $\gamma$, the herringbone angle $\theta$, inclination angle $\Theta$,  and torsion angle $\varphi_{C-C}$ are defined in Fig.~\ref{fig:illustrations}. $\rho$ is the mass density of the molecules.}
{\footnotesize %
\centering
\setlength{\tabcolsep}{0.125cm}
{\renewcommand{\arraystretch}{1.5}
\begin{tabular}{cccccccccc}
\hline
\hline 
                                       & $a$ [nm]           &  $b$ [nm]      &  $\gamma$ [\textdegree] &  $\theta$ [\textdegree] &  $\Theta$ [\textdegree] & $\varphi_{C-C}$ [\textdegree] & $\rho$ [g/cm$^3$] \tabularnewline
\hline 

\textit{p}-6P 2ML                           &   0.86$\pm$0.06  & 0.56$\pm$0.03   & 91.50$\pm$6.30    &   57.30$\pm$14.9           &   38.00$\pm$22.00      & 13.80$\pm$9.70     &  1.33$\pm$0.06\tabularnewline 

\textit{p}-6P4F 2ML                           &   0.97$\pm$0.08  & 0.51$\pm$0.06   & 81.60$\pm$18.5    &   62.00$\pm$18.50        &   38.00$\pm$22.00      & 17.00$\pm$10.70    &  1.31$\pm$0.09\tabularnewline

Ref. \cite{doi:10.1021/cg500234r}     & 0.827$\pm$0.01    & 0.548$\pm$0.01 &   89.80$\pm$3.30    & 61.7$\pm$13.70             &  17.70$\pm$6.00          & 15.70$\pm$7.90    &  1.29$\pm$0.02\tabularnewline  

Ref. \cite{BAKER19931571}              & 0.809            & 0.557           &    90.00         & 66.00                     &   18.00                    & 20.00              &  1.30\tabularnewline 
\hline
\hline 
\label{unitcell2}
\end{tabular}
}}
\end{table*}

In the case of the \textit{p}-6P 1ML and 2ML structures, the 1ML and 2ML unit-cells are rotated around the $z$-axis relative to each other by 111\textdegree.
The distance between the 1st and 2nd ML in $z$-direction is 2.431$\pm$0.262 nm.
In the case of the \textit{p}-6P4F 1ML and 2ML structures, the 1st and the 2nd ML unit-cells are rotated around $z$-axis relative to each other by 152\textdegree.
The distance between the 1st and the 2nd ML in $z$-direction is 2.473$\pm$0.405 nm.

Due to the repulsive interactions between the fluorinated molecular termini in \textit{p}-6P4F, the \textit{p}-6P4F structure shows a lower level of structural ordering compared to the \textit{p}-6P
(see Fig.~\ref{2ML_dens} d-f). The 1ML structure in both cases is less ordered compared to the 2ML structure, which comes as an effect of the interaction with the underlying silica surface.
Even though the molecules are packed in the bulk crystal structure, the repulsive molecule-molecule and attractive molecule-silica interactions decrease the degree of ordering
in the case of the \textit{p}-6P4F.

\subsection{Influence of the deposition rate on growth dynamics}

\label{dep_rate}
In this chapter we discuss the influence of the attempted deposition rates on the growth dynamics of the molecules. 
The deposition process was simulated with three different deposition rates: 3000$^{-1}$ ps$^{-1}$, 300$^{-1}$ ps$^{-1}$ and 30$^{-1}$ ps$^{-1}$. 
Results obtained by depositing molecules every 30 ps do not result in uniform, layer-by-layer growth, as observed for lower deposition rates. 
The underlying reason for this is that the timescale of 30 ps is between 1 and 2 orders of magnitude lower than the observed diffusion time scale. 
This prevents the molecule, once it is deposited, from diffusing to one of the neighbouring cluster units or ascending over the terrace to integrate into the existing layer, before the next molecule is deposited. 
This results in the formation of many irregular cluster shapes on the substrate with a surface density of $n$=3.04 nm$^{-2}$ for both \textit{p}-6P and \textit{p}-6P4F.
Fig.~4. of the Supporting Information compares the results for the average inclination angle, monolayer height and roughness between the deposition rates of 3000$^{-1}$ ps$^{-1}$ and 300$^{-1}$ ps$^{-1}$ at $T$ = 575 K, in case of the \textit{p}-6P.

Furthermore, as the deposition rate decreases from 300$^{-1}$ ps$^{-1}$  to 3000$^{-1}$ ps$^{-1}$, the critical surface density necessary for the molecules to have an orientational change
from a lying to an upright standing configuration decreases: Decreasing the rate by one order of magnitude results in the critical surface density
decrease from about 350 to about 200 molecules (in this case we define the critical cluster as the cluster comprising upright standing molecules). 
Lower deposition rates enable proper system equilibration and allow molecules to diffuse to nearby cluster
units and bind to them. 
Once the clusters become big enough, the number of upright standing molecules increases until it becomes energetically favorable for the other molecules to stand up as well.

In case of the deposition rate of 3000$^{-1}$ ps$^{-1}$  (see Fig.~\ref{3000ps} a)-d)), after the surface density of $n$=1.52 nm$^{-2}$ and $n$=1.75 nm$^{-2}$ for the \textit{p}-6P and \textit{p}-6P4F, respectively, is reached, they start forming stable, upright standing molecular clusters.
In case of the deposition rate of 300$^{-1}$ ps$^{-1}$, larger surface densities are required to observe the orientational change ($n$=3.04 nm$^{-2}$ for the \textit{p}-6P and $n$=2.66 nm$^{-2}$ for the \textit{p}-6P4F).
In case of the \textit{p}-6P the existence of bimodal phases (phases containing both lying and upright standing molecules) is observed after $n$=3.80 nm$^{-2}$. 
As the deposition continues and reaches $n$=4.18 nm$^{-2}$ for the \textit{p}-6P and $n$=3.88 nm$^{-2}$ for the \textit{p}-6P4F, all molecules (besides the wetting layer which is about 6\% and 4\% of the first monolayer (1ML) in case of \textit{p}-6P and \textit{p}-6P4F, respectively) have the upright standing orientation, with an average inclination angle of about
54\textdegree~and 68\textdegree. The value of 68\textdegree~ is comparable to the average inclination angle of pentacene molecules in a bulk-like monolayer adsorbed on {\it a}-SiO$_2$ 
calculated in \cite{doi:10.1002/cphc.200900084}, which is the stable low-temperature polymorph of crystalline pentacene.

\subsection{Conclusions}
In summary, we have theoretically investigated the role of change in polarity on nucleation and growth of the \textit{p}-6P and its symmetrically fluorinated derivative, \textit{p}-6P4F, on the amorphous silica surface.
We simulated the experimental vapor deposition process with all-atom molecular dynamics simulations, using three different deposition rates.
The growth of up to three complete monolayers is reproduced and monitored with the help of observables such as average height, surface roughness and average inclination angle as a function of time, together with the visual inspection of the grown structures.
In both cases, we observe the orientational change from lying to upright standing configuration once a critical surface density is reached.
During the formation of both the first and second monolayer, molecules tend to grow in chiral, fan-like, structures, where each consecutive molecule has a higher angle because it is propped by all the other molecules lying underneath. 
The growth of chiral islands is the main mechanism with which the growth of the \textit{p}-6P4F proceeds in the third layer, while the \textit{p}-6P, due to the lower interaction with the underlying substrate,
grows into islands with a lower degree of chirality. This ultimately leads to a lower barrier for step-edge crossing for the \textit{p}-6P4F, as the molecules adsorbed on the top of the chiral, structurally diverse island, have more different pathways available to ascend the terrace than the \textit{p}-6P on the \textit{p}-6P island.
Thus, \textit{p}-6P4F molecules have a relatively high chance to find an energy pathway low enough for them to descend the step-edge 
and integrate into the existing layer. 
We also measure the room-temperature unit-cell parameters in the first two layers. The values are well within the spread of the literature values, which confirm that both molecules grow into a single crystalline phase. 

An important implication of our findings is that partial fluorination of the \textit{p}-6P molecule can significantly alter its growth behaviour by modifying the rough, 3D growth into a smooth,
layer-by-layer growth in case of the \textit{p}-6P4F.
This has implications for the rational design of molecules and their functionalized forms which could be tailored for a desired growth behavior and HIOS structure formation.

\section{Acknowledgement}
The study was funded by the Deutsche Forschungsgemeinschaft (DFG, German Research Foundation) under Project No. 182087777-SFB 951 (project A7). 
The authors wish to thank Stefan Kowarik and Anton Zykov for inspiring discussions.

\newpage
\bibliography{literature}

\end{document}